\date{}
\documentclass[10pt,aps,prb,superscriptaddress,showpacs,twocolumn]{revtex4-1}
\usepackage{graphicx,psfrag,amsmath,amssymb,amsfonts,latexsym,color,dcolumn,bm}
\usepackage[hidelinks]{hyperref}

\begin{document}

\title{Photonic Spin Hall effect in bilayer graphene Moir\'e superlattices}

\author{W. J. M. Kort-Kamp}
\email{kortkamp@lanl.gov}
\affiliation{Theoretical Division and Center for Nonlinear Studies, Los Alamos National Laboratory, MS B262, Los Alamos, New Mexico 87545, USA}
\author{F. J. Culchac}
\affiliation{Instituto de F\'isica, Universidade Federal do Rio de Janeiro, Caixa Postal 68528, Rio de Janeiro 21941-972, RJ, Brazil}
\author{Rodrigo B. Capaz}
\affiliation{Instituto de F\'isica, Universidade Federal do Rio de Janeiro, Caixa Postal 68528, Rio de Janeiro 21941-972, RJ, Brazil}
\author{Felipe A. Pinheiro}
\affiliation{Instituto de F\'isica, Universidade Federal do Rio de Janeiro, Caixa Postal 68528, Rio de Janeiro 21941-972, RJ, Brazil}

\begin{abstract}
The formation of a superstructure - with a related Moir\'e pattern - plays a crucial role in the extraordinary optical and electronic properties of twisted bilayer graphene, including the recently observed unconventional superconductivity. Here we put forward a novel, interdisciplinary approach to determine the Moir\'e angle in twisted bilayer graphene based on the photonic spin Hall effect. We show that the photonic spin Hall effect exhibits clear fingerprints of the underlying Moir\'e pattern, and the associated light beam shifts are well beyond current experimental sensitivities in the near-infrared  and visible ranges. By discovering the dependence of the frequency position of the maximal photonic spin Hall effect shift on the Moir\'e angle, we argue that the latter could be unequivocally accessed via all-optical far-field measurements.  We also disclose that, when combined with the Goos-H\"anchen effect, the spin Hall effect of light enables the complete determination of the electronic conductivity of the bilayer. Altogether our findings demonstrate that sub-wavelength spin-orbit interactions of light
provide a unprecedented toolset for investigating optoelectronic properties of multilayer two-dimensional van der Waals materials.  
\end{abstract}
\maketitle

 At macroscopic scales geometric optics provides an adequate description of several photonic phenomena. However, in the sub-wavelength regime light's spatial and polarization degrees of freedom are not independent quantities, resulting in deviations from the traditional ray optics picture due to optical spin-orbit interactions \cite{Aiello2015, Bliokh2015, Bliokh20152, Ling2017, Aiello2014, Aiello2015PRL,Onoda2004,Bliokh2006,Hosten2008, Luo2011Brewster}. The scattered electromagnetic field due to a linearly polarized finite width beam impinging on an interface develops a non-trivial spin texture arising from the shift of photons with contrary helicity to opposite edges of the beam cross section, a phenomenon known as the spin Hall effect of light (SHEL) \cite{Onoda2004, Bliokh2006, Hosten2008,Ling2017}.  The SHEL is ubiquitous to any interface and represents a remarkable failure of geometric optics at the nanoscale. Recent experimental demonstrations have shown that the SHEL is uniquely suited for applications in precision metrology, including nanoprobing \cite{Herrera2010}, thin films characterization \cite{Zhou20122, Qiu2014}, mapping of absorption mechanisms in bulk semiconductors \cite{Menard2009, Menard2010}, and multilayer graphene identification in the absence of interlayer coupling \cite{Zhou2012}. More recently, it has been proposed that the SHEL can be used to probe the quantum Hall effect in monolayer graphene \cite{KortKamp2016, Cai2017}, topological phase transitions in the expanded graphene family \cite{KortKamp2017}, and hyperbolic behavior in anisotropic 2D atomic crystals \cite{Zhang2018}.  

In recent years, another class of 2D materials has drawn a lot attention, namely van der Waals structures assembled together layer-by-layer with controllable sequence and orientation~\cite{geim2013}. These structures exhibit unusual physical properties that cannot be found in either monolayers or in bulk.  Twisted Bilayer Graphene (TBG) is one important example of these multilayer materials, where the twist produces a Moir\'e pattern and induces a static periodic potential from the coupling between graphene layers, leading to an angle-dependent band structure~\cite{santos2007}. TBG can occur naturally by chemical vapor deposition\cite{li2009,lu2013,nie2011} or fabricated by mechanically folding graphene~\cite{carozo2013} and stacking two monolayers together~\cite{kim2016}. The band structure of TBG has been investigate via different approaches\cite{Ni2008,Righi2011,Sato2012,Robinson2013,Wang2010,Ohta2012,Brihuega2012,Mikito, luican2011,wu2012, yin2015, cao2018,cao2018b} which  lead to the discovery of remarkable electronic properties,  including Dirac-like spectrum~\cite{luican2011}, low energy van Hove singularities~\cite{li2009,wu2012}, localization of low energy states~\cite{yin2015}, and unconventional superconductivity~\cite{cao2018,cao2018b}. This latter result is the first evidence of a purely carbon-based 2D superconductor, and occurs for some specific ``magic'' Moir\'e angles between the two graphene layers. At these angles TBG exhibits ultraflat bands near charge neutrality, leading to correlated insulating states. Upon electrostatic doping superconductor states exist for relatively high critical temperatures, with a phase diagram similar to that of cuprates. All these findings  show that the Moir\'e superlattice plays a pivotal role in the extraordinary optical, thermal, and electrical properties of TBG. Hence, the precise characterization of the Moir\'e pattern of such structures is crucial for the development of future TBG-based technologies.  

Here we put forward an all-optical approach, merging electronic properties of multilayer van der Waals materials and nanoscale photonic spin-orbit interactions, to determine the Moir\'e pattern in TBG via the photonic spin Hall effect. We demonstrate that SHEL shifts depend on the Moir\'e angle and are within current experimental capabilities for detecting beam shifts in the near-infrared and visible ranges. We show that near the Brewster angle the SHEL is greatly magnified as it scales with the inverse of the fine structure constant $\alpha$. This is an outstanding advantage over  previous works where the probed optical response of TBG is typically linear in $\alpha$, resulting in a weak signal that may be screened by spurious effects. We propose a simple model to describe the frequency position of the maximal SHEL shift as a function of the Moir\'e angle, allowing for its reliable determination. We also demonstrate that, when combined with the Goos-H\"anchen effect, the SHEL enables full far-field characterization of the electronic conductivity of the bilayer.   

\section{Mehods}

\begin{figure}
\centering
\includegraphics[width=\linewidth]{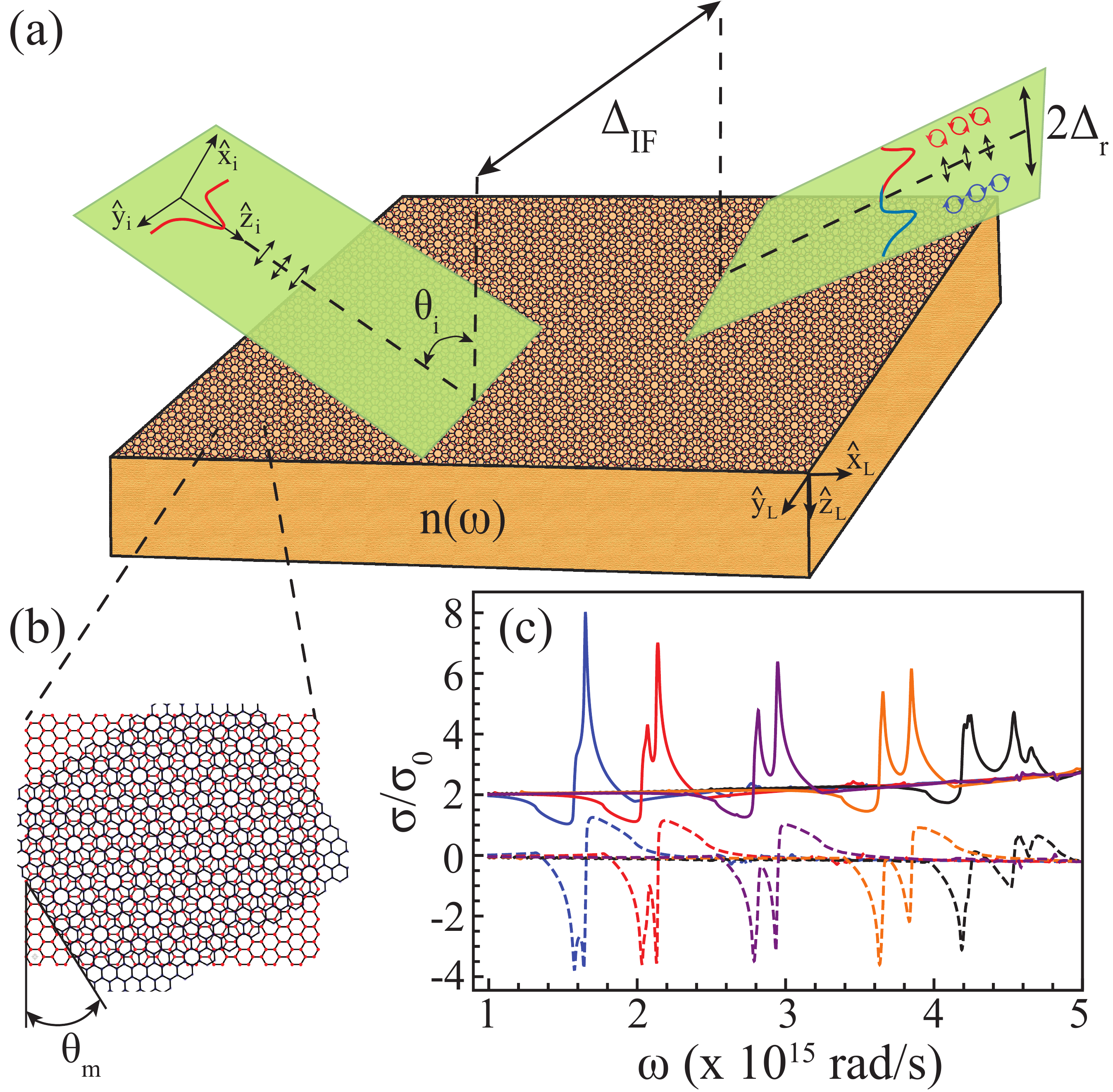}
\caption{Schematic representation of the system under study. (a)  Non-trivial spin structure in the reflected field due to a linearly polarized gaussian beam impinging on the interface air/bilayer graphene coated substrate. (b) Moir\'e pattern formed in bilayer graphene and definition of the Moir\'e angle $\theta_{\rm m}$ between the two monolayers composing the superlattice. (c) Real (solid) and imaginary (dashed) components of the electronic conductivity $\sigma(\omega)$ of neutral bilayer graphene versus frequency for $\theta_{\rm m} = 7.34^o$ (blue), $9.43^o$ (red), $21.8^o$ (black), $42.1^o$ (orange), and $46.83^o$ (purple). Here  $\sigma_0 = e^2/4\hbar$ is the universal conductivity of monolayer graphene.} 
\label{Fig1}
\end{figure}

Let us consider the system depicted in Fig. \ref{Fig1}a. A gaussian beam of frequency $\omega$ propagating in vacuum and confined along the $y$-direction impinges at an angle $\theta_{\rm i}$ on a flat bilayer graphene superlattice which is on top of a substrate of refractive index $n(\omega)$. The incident electric field can be cast as 
\begin{equation}
 {\bf E}_{i}= \mathcal{A}(y_i,z_i) [f_p \hat{{\bf x}}_i + f_s \hat{{\bf y}}_i - i f_s  k_0 y_i (\Phi+ik_0 z_i)^{-1} \hat{{\bf z}}_i]\, , 
\end{equation}
where $\mathcal{A}(y,z)= \sqrt[4]{\frac{2}{\pi w_0^2 (1+k_0^2 z^2/\Phi^2)}} e^{i k_0 z - \frac{k_0^2y^2}{2(\Phi+ik_0 z)}}$ is the Gaussian amplitude, $k_0=\omega/c$ is the wavenumber, $w_0$ is the beam waist, and $\Phi=k_0^2w_0^2/2$ is the Rayleigh range~\cite{BornWolf}. The  unit vectors ($\hat{\bf x}_i, \hat{\bf y}_i,\hat{\bf z}_i$) are associated to a reference frame of coordinates ($x_i,y_i,z_i$) attached to the central plane wave component of the incident beam with the origin at the point where the latter reaches the surface. The polarization state of the incident beam is characterized by the complex unit vector $\hat{{\bf f}}= f_{p}{\bf\hat{x}}_{i}+f_{s}e^{i\eta}{\bf\hat{y}}_{i}$, where $f_p$ and $f_s$ are real valued amplitudes related to the  vertical and horizontal polarization components and $\eta$ is the relative phase shift between them. In the following, we will compute the SHEL of the reflected field ${\bf E}_r$.

A simple expression for ${\bf E}_r$ can be derived by enforcing the reflection law and Fresnel's equations for each plane wave component of the incident beam, as well as using the paraxial approximation. In this case \cite{KortKamp2016, Cai2017, KortKamp2017},
\begin{eqnarray}
\!\!\! \!\!\! {\bf E}_r \!&\propto&\!  \left[(1\!+\!i\rho_R)\mathcal{A}(y_r\!-\!\tilde{\delta}_-,z_r) - \rho_I\mathcal{A}(y_r\!-\!\tilde{\delta}_+,z_r) \right] \hat{{\bf e}}_{-}  \cr
\!\!\!\!\!\!\!\!\!   &+& \!  \left[(1\!-\!i\rho_R)\mathcal{A}(y_r\!-\!\tilde{\delta}_+,z_r) + \rho_I\mathcal{A}(y_r\!-\!\tilde{\delta}_-,z_r) \right] \hat{{\bf e}}_{+}  .
\label{FieldSHEL}
\end{eqnarray}
Here, $\rho = \rho_R + i\rho_I =f_s e^{i\eta}r_s/f_pr_p$, where $r_{s,p} = R_{s,p}e^{i\phi_{s,p}}$ are the complex reflection coefficients for the central plane wave component, namely \cite{NunoBook}
\begin{eqnarray}
r_s &=& \frac{\cos\theta_i-\sqrt{n^2-\sin^2\theta_i}-Z_0\sigma}{\cos\theta_i+\sqrt{n^2-\sin^2\theta_i}+Z_0\sigma} \, ,\cr
r_p&=& \frac{n^2\cos\theta_i-\sqrt{n^2-\sin^2\theta_i}(1-Z_0\sigma\cos\theta_i)}{n^2\cos\theta_i+\sqrt{n^2-\sin^2\theta_i}(1+Z_0\sigma\cos\theta_i)}\, , 
\end{eqnarray}
where $Z_0 = \sqrt{\mu_0/\epsilon_0}$ is the vacuum impedance and $\sigma(\omega)$ is the bilayer's conductivity. The reference frame of coordinates $(x_r, y_r, z_r)$ has the same origin as $(x_i, y_i, z_i)$ and unit vectors ${\bf\hat{x}}_r={\bf\hat{x}}_i - 2 {\bf\hat{x}}_L ({\bf\hat{x}}_i \cdot {\bf\hat{x}}_L)$,  ${\bf\hat{y}}_r={\bf\hat{y}}_i$, and ${\bf\hat{z}}_r={\bf\hat{z}}_i - 2 {\bf\hat{z}}_L ({\bf\hat{z}}_i \cdot {\bf\hat{z}}_L)$, where ($\hat{\bf x}_L,\hat{\bf y}_L,\hat{\bf z}_L$) are laboratory frame versors (see Fig. \ref{Fig1}).  Also,  $\hat{{\bf e}}_{\pm} = [\hat{{\bf x}}_r\pm i(\hat{{\bf y}}_r-\beta y_r\hat{{\bf z}}_r)]/\sqrt{2}$ are left and right ``circularly" polarized unit vectors with $\beta = ik_0/(\Phi+ik_0z_r)$.  The complex displacements $\tilde{\delta}_{\pm}$ are given by  $ \tilde{\delta}_{\pm} = w_pY_p (1\pm i\rho)+w_sY_s(1\mp ie^{-i\eta}/\rho^*)\, ,  $ where $w_{s,p} = f_{s,p}^2 R_{s,p}^2/[f_p^2R_p^2+f_s^2R_s^2]$ give the amount of energy stored in $s$ and $p$ polarizations, and $Y_p = if_se^{i\eta}\cot\theta_i(1+r_s/r_p)/f_pk_0 = -\left.Y_s\right|_{p\leftrightarrow s, \eta \rightarrow 0}$.   The reflected beam in Eq. (\ref{FieldSHEL}) is given by the superposition of left and right circularly polarized states, each component corresponding to the weighted average of two gaussians centered at $\tilde{\delta}_-$ and $\tilde{\delta}_+$. The well known results for the SHEL reported in \cite{Bliokh2006, Hosten2008} are recovered in the limit of low-dissipative  systems ($\rho_I \ll \rho_R$). In this case, each polarization state is reduced to a single gaussian and photons of opposite helicity are shifted by Re($\tilde{\delta}_-$) (right circular polarization) and Re($\tilde{\delta}_+$) (left circular polarization). 

The SHEL displacement for the $\hat{{\bf e}}_{-}$ and $\hat{{\bf e}}_{+}$ components, calculated via the intensity distribution centroid, can be conveniently expressed in terms of the spatial Imbert-Fedorov shift  $\Delta_{\rm IF}$ of the whole reflected beam \cite{Fedorov1955,Imbert1972} and of the relative shift $\Delta_{\rm r}$ between the right and left polarizations,  ${\Delta}^{\pm}_{\rm SHEL} = \Delta_{\rm IF}  \pm \Delta_{\rm r}$ (see Fig. \ref{Fig1}),
\begin{eqnarray}
\label{DeltaIF}
\Delta_{\rm IF}\!\! &=&\!\frac{\cot\! \theta_{\rm i}}{k_0}\!\!\left[\!\sqrt{w_sw_p} \sin(\!\phi_p\!\!-\!\phi_s\!\!-\!\eta\!)\!\!-\!\!\frac{f_s^2w_p^2\!+\!\!f_p^2w_s^2}{f_sf_p}\!\sin\eta\right],  \cr
\label{Deltar}
\Delta_{\rm r}\ &=&\frac{\cot\! \theta_{\rm i}}{k_0}\!\!\left[1+\frac{\sqrt{w_sw_p}}{f_sf_p}\cos(\phi_p-\phi_s)\right]\, .
\end{eqnarray}
%

\begin{figure*}
\centering
\includegraphics[width=0.95\linewidth]{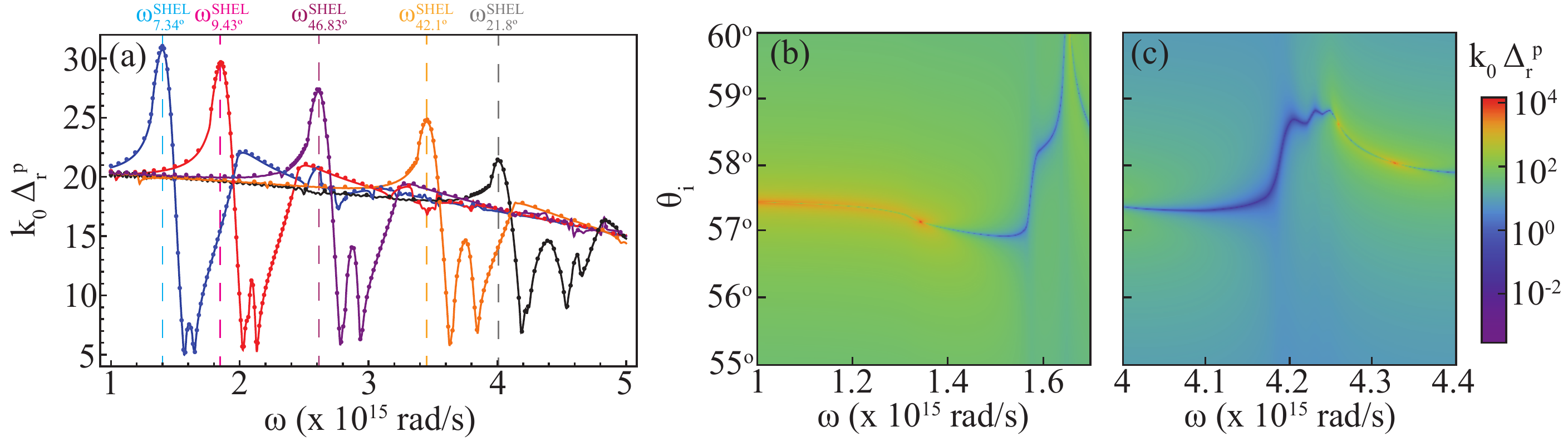}
\caption{Spin Hall effect of light in bilayer graphene. (a) Relative SHEL shift versus frequency for a $p$-polarized gaussian beam impinging on the system at $\theta_{\rm i} \simeq 56.3^o$, corresponding to the Brewster angle of the substrate. The dotted curves were computed using the approximated results in Eq. (\ref{DeltarApprox}), showing an excellent agreement with  the full numerical calculations  (solid curves).The vertical dashed lines mark the frequency position $\omega^{\rm SHEL}_{\theta_{\rm m}}$ of the maximum SHEL shift for neutral bilayer graphene. All other parameters are the same as in Fig. \ref{Fig1}. $\Delta_{\rm r}$ dependence with the incidence angle around the resonant frequencies of neutral bilayer graphene for (b) $\theta_{\rm m} \simeq 7.34^o$ and (c) $\theta_{\rm m} \simeq 21.8^o$.} 
\label{Fig2}
\end{figure*}

To numerically evaluate the SHEL shift due to bilayer graphene, one needs to compute its conductivity $\sigma({\omega})$.   The optoelectronic response of bilayer graphene is modeled using a tight-binding hamiltonian based on a linear combination of orthogonal $p_z$ atomic orbitals \cite{Mikito, RevModPhys2009}, 
$ \hat{H}=-\sum_{i,j}t({\bf R}_i-{\bf R}_j)|{\bf R}_i\rangle \langle {\bf R}_j |+ h.c. \, ,  $ where  ${\bf R}_i$ and $|{\bf R}_i\rangle $ are the position vector and atomic state at site $i$. The hopping energy $t ({\bf R}_i -{\bf R}_j )$ between orbitals $i$ and $j$ depends on interatomic distances and relative orientation between $p_z$ atomic orbitals at each site \cite{Mikito}, 
\begin{equation}
t({\bf d})=V_{pp\pi}(d)\left[\frac{d_z^2}{d^2} -1\right]- V_{pp\sigma}(d)\frac{d_z^2}{d^2}\, , 
\end{equation}
where  $d = |{\bf d}|=|{\bf R}_i-{\bf R}_j|$, and the interatomic distance dependent hopping energies are  $V_{pp\pi}(d)=V_{pp\pi}^0 e^{-(d-a_0)/\delta_0}$ and $V_{pp\sigma}(d)=V_{pp\sigma}^0 e^{-(d-d_0)/\delta_0}$. Here, $a_0 = 0.142$ nm is the intralayer nearest neighbor distance, $d_0 =  0.335$ nm is the interlayer spacing, and $\delta_0 = 0.317a_0$ is the decay length of the hopping integral, chosen so that the next-nearest neighbor hopping in the monolayer becomes $0.1V_{pp\pi}^0$. Also, in order to fit the dispersions of monolayer graphene and AB-stacked bilayer graphene one sets $V_{pp\pi}^0=-2.7$ eV and $V_{pp\sigma}^0 = -0.48$ eV  \cite{Mikito}. In our calculations we consider only incident transverse magnetic radiation (electric field along $x_i$ axis) and, hence, only the diagonal component of the 2D conductivity in the $x_{\rm L}$ direction needs to be computed. Similarly to previous works on the optoelectronic properties of bilayer graphene \cite{Mikito,RevModPhys2009,Moon2015_2, Vela2018}, the local longitudinal conductivity $\sigma(\omega)= \sigma_{\rm intra}(\omega)+\sigma_{\rm inter}(\omega)$ is determined by applying well known Kubo's linear response formalism \cite{Kubo-I-1957, Ventura,Lopez2018}, where the first (second) term accounts for intraband (interband) processes. In the following we consider only the case of neutral TBG, resulting in $\sigma_{\rm intra}(\omega) = 0$.  
The interband component of the conductivity is computed via 
\begin{equation}
\frac{\sigma_{\rm inter}(\omega)}{\sigma_0}\!=\! -\frac{4i}{S}\!\!\sum_{\stackrel{\mbox{\small {\bf k}}}{\lambda \neq \lambda'}}\!\!
\frac{f_{{\bf k}\lambda}\!\!-\!\!f_{{\bf  k} \lambda'}}{\epsilon_{{\bf k} \lambda}\!-\! \epsilon_{{\bf  k} \lambda'}}
\frac{ |\! \left\langle {\bf k}\lambda | \hbar \hat{v}_x |{\bf k} \lambda' \right\rangle \! |^2}{\epsilon_{{\bf  k} \lambda}\! -\! \epsilon_{{\bf  k} \lambda'}\!+\!\hbar (\omega\!+\!i\gamma)},
\end{equation}
where  $\sigma_0 = e^2/4\hbar = \pi\alpha/Z_0$ is the fundamental conductance of monolayer graphene and $\alpha$ is the fine structure constant. Here, $\gamma$ is a phenomenological broadening energy,  which is necessary to avoid singularities in the numerical calculations. The actual value of $\gamma$ strongly depends on the quality of the bilayer graphene sample in consideration. We choose $\hbar\gamma = 3$ meV\cite{Vela2018} in order to potentially distinguish absorption peaks that emerge in the electromagnetic field's spectra due to electron-hole asymmetry for Moir\'e angles $\gtrsim10^o$.  Larger values of $\gamma$ will result in bandwidth broadening and reduced spectral resolution of resonant absorption peaks. Also, $S$  is the area of the system, $\epsilon_{{\bf k}\lambda}$ are the eigenenergies of the tight-binding hamiltonian for charge carries at band $\lambda$ with momentum $\hbar{\bf k}$, $ |{\bf k}\lambda\rangle $ are the associated eigenstates, $f_{{\bf k}\lambda} = [1+e^{\epsilon_{{\bf  k} \lambda}/k_BT}]^{-1}$ is the Fermi-Dirac distribution function at temperature $T$ and zero doping, and  $\hat{v}_x= d\hat{x}/dt = [\hat{x},\hat{H}]/i\hbar$ is the $x$-component of the velocity operator \cite{Brey}. 

\section{Results and Discussion}

Let us discuss the optoelectronic properties of bilayer graphene in order to investigate their fingerprints into the photonic spin Hall effect. In Fig. \ref{Fig1}b we show the definition of the Moir\'e angle $\theta_{\rm m}$ characterizing the relative twist between the two graphene monolayers that compose the superlattice. The interlayer coupling in bilayer graphene gives rise to a band structure (not shown) that is significantly distinct from that of two non-interacting graphene monolayers, specially near to band anticrossing regions that result in van Hove singularities in the electronic local density of states \cite{RevModPhys2009,Mikito}. These singularities systematically change with the rotation angle $\theta_{\rm m}$, and one expects the optical response of bilayer graphene to be similarly affected by the Moir\'e pattern formed in the superlattice. By using the tight-binding model and the formalism described above we have computed the longitudinal conductivity as a function of frequency for various Moir\'e patterns. Fig. \ref{Fig1}c shows the zero-temperature real $\sigma_R(\omega)$ (solid curves) and imaginary $\sigma_I(\omega)$ (dashed curves) components of the conductivity in the near-infrared to visible frequency range for neutral bilayer graphene. The conductivity is real valued and  approaches $2\sigma_0$ in the low-frequency limit, corresponding to the case of two uncoupled graphene monolayers. However, in a narrow frequency range both $\sigma_R(\omega)$ and $\sigma_I(\omega)$ feature a highly non-monotonic behavior, as it can be seen for instance in the curve  for $\theta_{\rm m} = 7.34^o$ around $\omega \simeq 1.6 \times 10^{15}$ rad/s. The double-peak structure present in the conductivity,  more easily seen at large Moir\'e angles, can be traced back to the electron-hole asymmetry and interband excitations between van Hove singularities arising from band-anticrossing of twisted Dirac cones \cite{Mikito, Moon2015_2}. The most remarkable aspect of Fig. \ref{Fig1}c for our work is the fact that the resonant behavior in $\sigma(\omega)$ shifts from the near-infrared to visible wavelengths as one increases the rotation angle between the monolayers from $0^o$ to $30^o$. The opposite trend is observed as $\theta_{\rm m}$ changes from $30^o$ to $60^o$, since a superlattice with rotation angle  $\theta_{\rm m}$  is equivalent to a bilayer graphene with a twist angle $60^o-\theta_{\rm m}$ followed by a relative translation of one of the layers by one period. 
\begin{figure}
\centering
\includegraphics[width=\linewidth]{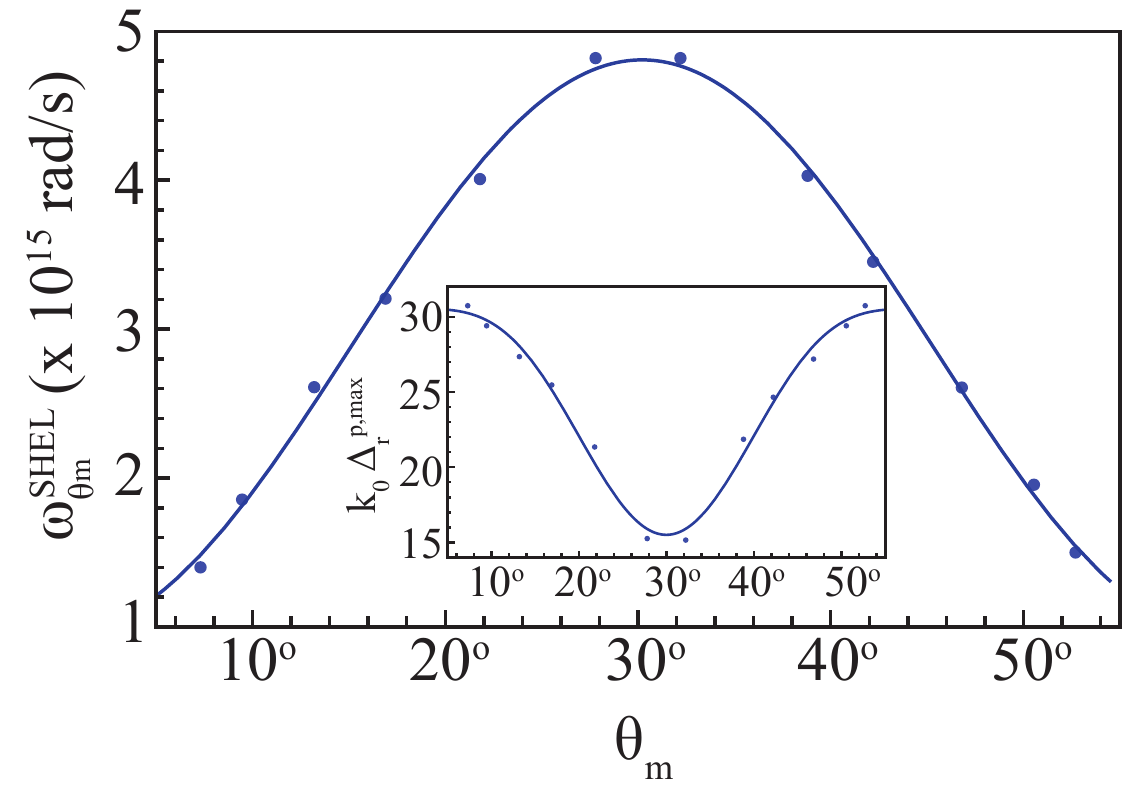}
\caption{Identifying Moir\'e superlattices with the spin Hall effect of light. Frequency $\omega_{\theta_{\rm m}}^{\rm SHEL}$ associated to the maximum SHEL shift in neutral bilayer graphene at various Moir\'e angles for $p$-polarization and $\theta_{\rm i} \simeq 56.3^o$ (dots). The solid line is the resulting interpolation of $f(\theta_{\rm m}) = A + B\cos^2(C \theta_{\rm m})$ to the numerical data, where $A \simeq 9.5 \times 10^{14}$ rad/s, $B \simeq 3.85\times 10^{15}$ rad/s, and $C \simeq 2.98$.  The inset shows $k_0 \Delta_{\rm r}$  versus $\theta_{\rm m}$ (dots) at $\omega  = \omega_{\theta_{\rm m}}^{\rm SHEL}$  and the corresponding  interpolated curve $g(\theta_{\rm m}) = A + B\cos^4(C \theta_{\rm m})$, where  $A \simeq 30.54$, $B \simeq -15.04$, and $C \simeq 3$. } 
\label{Fig3}
\end{figure}

We now turn our attention to the photonic spin Hall effect. For incident $s$ or $p$ linearly polarized  gaussian beams, the spatial Imbert-Fedorov shift vanishes as a consequence of the mirror symmetry with respect to the incidence plane. In this case, left and right circularly polarized photons are shifted to positive and negative values along the $y-$direction while keeping the intensity centroid of the whole reflected beam confined to the $x_Lz_L$ plane. In Fig. \ref{Fig2}a we show $\Delta_{\rm r}$ versus frequency for a $p$-polarized incident beam and a few representative Moir\'e angles of the superlattice.  The bilayer lies on top of a transparent substrate of constant refractive index $n(\omega) = 1.5$ in the wavelength range of interest.  In order to magnify the impact of interlayer coupling in the dispersion curves of the SHEL we choose $\theta_{\rm i} = 56.3^o$, since this corresponds to the Brewster angle of the air-substrate interface and $\Delta_{\rm r} \propto R_s/R_p  \gg 1 $ for a purely $p$-polarized gaussian wave packet \cite{Luo2011Brewster} [see Eq. (\ref{Deltar})]. For a fixed $\theta_{\rm m}$ the resulting relative shift is a quasi-flat function of $\omega$ ($k_0\Delta_{\rm r}\simeq 20$) for most of the spectral range considered, since both $R_{s,p}$ and $\phi_{s, p}$ are constants.  A different landscape occurs around the conductivity's resonant frequency: the increased absorption in the bilayer results in changes in the reflection amplitude as well as in a non-trivial dispersion curve for the phase of the reflected field. Consequently,  $\Delta_{\rm r}$ remarkably differs from the constant background shift described above, and features a peak followed by a steep suppression of the SHEL as one crosses the resonance region  increasing $\omega$. The variation between the maximum and the two nearby minima values of $k_0\Delta_{\rm r}$ can be as high as $\sim\! 25$ for small Moir\'e angles.  The SHEL peak consistently moves to higher frequencies as one increases $\theta_{\rm m}$ up to $30^o$, and then hops to lower frequencies in the range $30^o < \theta_{\rm m} <60^o$, similarly to the changes in $\sigma(\omega)$ due to van Hove singularities. Qualitatively similar plots can be obtained for other polarization states of the incident light, but a weaker influence of van Hove singularities in the SHEL spectrum is observed.

A simple expression describing $\Delta_{\rm r}$ for incidence at the substrate's Brewster angle can be derived by performing an expansion of Eq. (\ref{Deltar}) in powers of $Z_0|\sigma(\omega)| \ll 1$, leading to
\begin{equation}
\label{DeltarApprox}
\Delta_{\rm r}^{\theta_{\rm i}= \theta_{\rm B}}\! \simeq\! -\frac{1}{nk_0}\!\left[\!1\!-\!\frac{n^2\!+\!3}{n^2\!+\!1}\!-\!\frac{2(n^2\!-\!1)}{Z_0\sqrt{n^2\!+\!1}}\frac{\sigma_R(\omega)}{|\sigma(\omega)|^2} \right]\, , 
\end{equation}
where we used that $\theta_{\rm i} = \theta_{\rm B} = \tan^{-1}n(\omega)$. Figure \ref{Fig2}a shows the plot of Eq. (\ref{DeltarApprox}) versus $\omega$ (dotted curves), which are in excellent agreement with the full numerical calculations following from Eq. (\ref{Deltar}) over the entire frequency range and for all Moir\'e angles investigated.  The first two terms in Eq. (\ref{DeltarApprox}) arise solely from the substrate and are constant due to the dispersionless refractive index we consider here.  The last term encodes all the information on interlayer coupling in the superlattice and typically dominates over the others as it is of the order of $(Z_0\sigma_0)^{-1} = (\pi\alpha)^{-1} \sim 40$. Note that the relative SHEL shift depends on both real and imaginary components of the electronic conductivity, $k_0\Delta_{\rm r} \propto \sigma_R/Z_0(\sigma_R^2+\sigma_I^2)$, which explains the intricate  behavior observed near resonances. This dependence contrasts with characterization techniques of 2D materials based on absorption measurements, which generally are sensitive to $\sigma_R(\omega)$ and linear in $\alpha$ to leading order. It is worth emphasizing that the scaling of the SHEL with the inverse of the conductivity results from choosing the incident angle to coincide with the Brewster angle of the substrate, which enhances the contrast between maxima and minima already present in the SHEL spectrum. This dependence with $\sigma^{-1}$ seems paradoxal at first glance, as it would result in an infinity shift in the limit $\sigma \rightarrow 0$ (absence of bilayer graphene). This paradox, however, is resolved by noticing that in the aforementioned limit the intensity of the reflected beam approaches zero for incident p-polarized light.

In Figures \ref{Fig2}b,c we show the SHEL relative shift dependence with the incidence angle of the gaussian beam for frequencies around the resonance in two different Moir\'e patterns, $\theta_{\rm m} = 7.34^o$ and $\theta_{\rm m} = 21.8^o$, respectively. Although the resonant/anti-resonant behavior can be observed for any incidence angle, a better contrast between the peak/minima induced by the interlayer coupling and the constant contribution arising from the substrate is more clearly observed in the range $55^o < \theta_{\rm i} < 60^o$. Beam shifts several orders of magnitude larger than the wavelength of incident light can be achieved in a narrow range of incident angles, which approach the true Brewster angle of the substrate/bilayer graphene system. Note also that both plots feature a contour structure that highlights the maxima and minima of $\Delta_{\rm r}$, being strongly dependent of the value of $\theta_{\rm m}$. We checked that these contours are different for all the Moir\'e patterns ($0^o < \theta_{\rm m}< 30^o$) considered in our calculations.  We also mention that although the reflectivity of the system (not shown)  in the range of incidence angles under consideration in Fig. \ref{Fig2}b,c is small (typically $\lesssim 1\%$), the SHEL shifts could be accessed within weak measurements framework via signal enhancement techniques, similarly to the experimental demonstration in Ref. [10] for light impinging on a substrate near and at the Brewster angle.

Our results suggest that we could take advantage  of the SHEL displacement dependence over $\theta_{\rm m}$ to characterize the Moir\'e pattern in a bilayer graphene sample. In Figure  \ref{Fig3} we show the  computed frequency  $\omega_{\theta_{\rm m}}^{\rm SHEL}$ (dots) of the resonant peak of the SHEL relative shift (marked in Fig. \ref{Fig2}a by vertical dashed lines) for various values of $\theta_{\rm m}$ and the same gaussian beam considered in Fig. \ref{Fig2}.  We have fitted our results for  $\omega_{\theta_{\rm m}}^{\rm SHEL}$ using the function $f(\theta_{\rm m}) = A + B\cos^2(C \theta_{\rm m})$, where $A,\ B,\ C$ are fitting parameters. An excellent  agreement with the numerical data is obtained for $A \simeq 9.5 \times 10^{14}$ rad/s, $B \simeq 3.85\times 10^{15}$ rad/s, and $C \simeq 2.98$, providing a powerful tool for determining the geometric pattern in bilayer graphene samples. Surprisingly, we checked that the value of the peak of $\Delta_{\rm r}$ also shows a harmonic behavior with $\theta_{\rm m}$, presenting a maximum near AA and AB stacking ($\theta_{\rm m} = 0^o, 60^o$, respectively) and a minimum as one approaches $\theta_{\rm m} = 30^o$. In this case the results for $k_0\Delta_{\rm r}$ are well fitted by $g(\theta_{\rm m}) = A + B\cos^4(C \theta_{\rm m})$, with fitting parameters  $A \simeq 30.54$, $B \simeq -15.04$, and $C \simeq 3$.
\begin{figure}
\centering
\includegraphics[width=\linewidth]{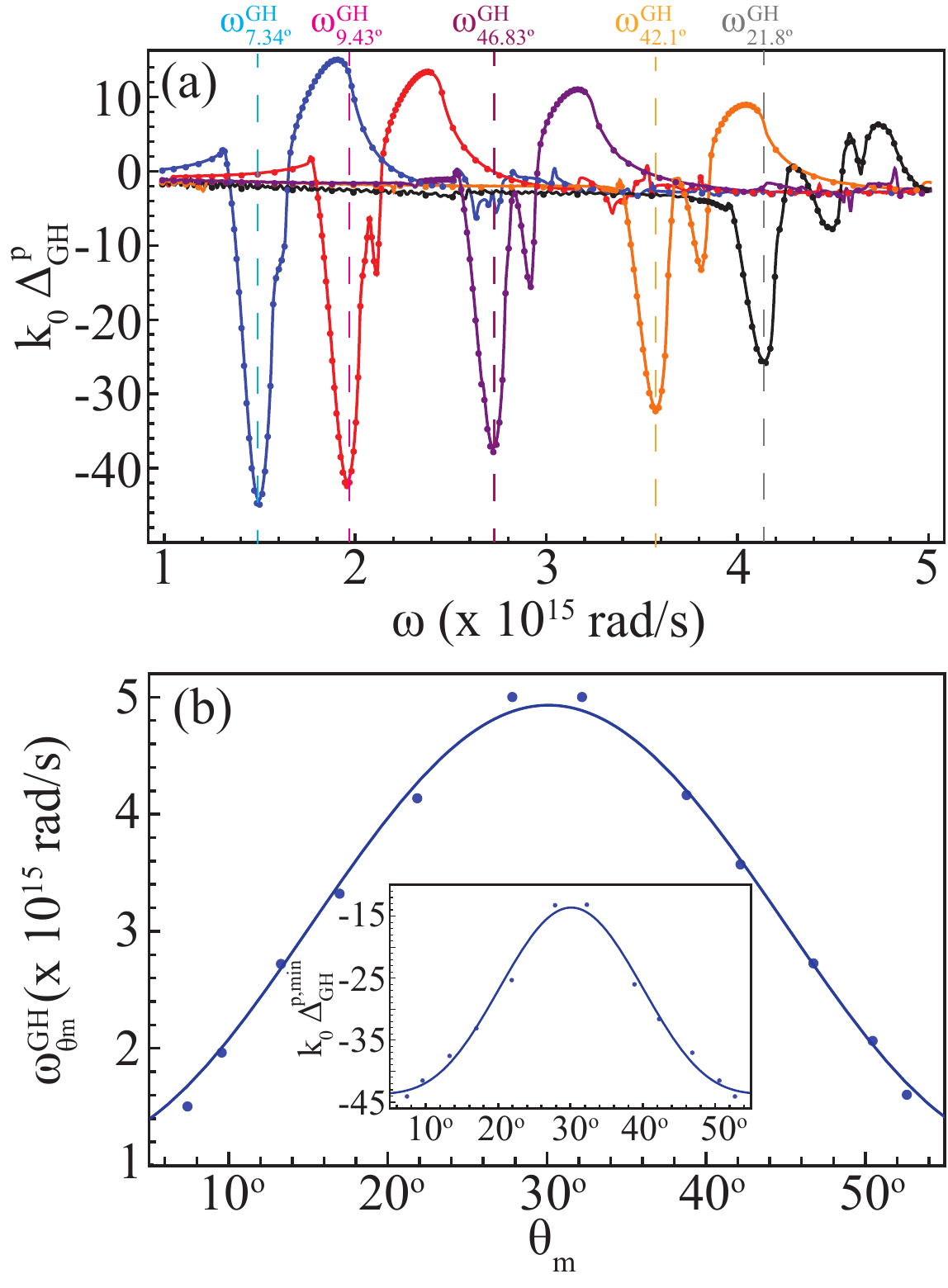}
\caption{Identifying Moir\'e superlattices using the Goos-H\"anchen effect. (a) Spatial Goos-H\"anchen beam shift versus frequency for the same parameters and Moir\'e angles as in Figs. \ref{Fig1} and \ref{Fig2}. The vertical dashed lines mark the frequency position $\omega^{\rm GH}_{\theta_{\rm m}}$ of the minimum (maximum in absolute value) GH shift in neutral bilayer graphene. (b) $\omega^{\rm GH}_{\theta_{\rm m}}$  versus the Moir\'e angle $\theta_{\rm m}$. Similarly to Fig. \ref{Fig3}, the interpolated curve is given by $f(\theta_{\rm m})$, where  $A \simeq 1.15 \times 10^{15}$ rad/s, $B \simeq 3.78\times 10^{15}$ rad/s, and $C \simeq 2.99$. The inset corresponds to $k_0 \Delta_{\rm GH}$  versus $\theta_{\rm m}$ (dots) for $\omega  = \omega_{\theta_{\rm m}}^{\rm GH}$. The numerical data was interpolated using $g(\theta_{\rm m})$, where  $A \simeq -43.78$, $B \simeq 30.07$, and $C \simeq 3$.
} 
\label{Fig4}
\end{figure}

We mention that in addition to the SHEL displacements  $\Delta_{\rm IF}$ and $\Delta_{\rm r}$, an angular Imbert-Fedorov beam shift can take place upon reflection of the gaussian beam at the interface between two media (not shown in Fig. \ref{Fig1} for clarity). However, it vanishes for incident linearly polarized light similarly to $\Delta_{\rm IF}$. When the incident gaussian beam is confined within the plane of incidence other effects can occur, namely the spatial and angular Goos-H\"anchen (GH) shifts \cite{Goos1947}. The angular GH shift is proportional to\cite{Grosche2015,KortKamp2016} $\partial_{\theta_{\rm i}}R_{s,p} \propto (Z_0\sigma_0)^2 = (\pi\alpha)^2$,  which prevents one to distinguish the resonances from the spurious background contribution due to the substrate (unlike the SHEL shift which is enhanced owing to the high $R_s/R_p \sim (\pi\alpha)^{-1}$ factor near the pseudo-Brewster angle). On the other hand, the spatial GH displacement, $k_0\Delta_{\rm GH} = w_s\partial_{\theta_{\rm i}} \phi_s+ w_p\partial_{\theta_{\rm i}} \phi_p$, allows for identifying the value of $\theta_{\rm m}$ in the superlattice, as the phase of  the reflected field typically presents strong variations near material resonances. In Figure \ref{Fig4}a we present the results for the spatial GH shift as a function of frequency for the same parameters as in Fig. \ref{Fig2}. Similarly to the relative SHEL shift, a strong variation in $\Delta_{\rm GH}$ can be observed near the resonances of the conductivity, with the GH shift changing from negative to positive values. The plot also shows that the approximated expression 
\begin{equation}
\Delta_{\rm GH}^{\theta_{\rm i}= \theta_{\rm B}}\! \simeq\! \frac{\sqrt{n^2+1}(n^4-1)}{Z_0n^3k_0} \frac{\sigma_I(\omega)}{|\sigma(\omega)|^2}\, ,
\label{DeltaGHApprox}
\end{equation}
perfectly describes the full numerical calculations. We show in Fig. \ref{Fig4}b how the frequency $\omega^{\rm GH}_{\theta_{\rm m}}$ corresponding to the position of the minimum in each curve in Fig. \ref{Fig4}a (as indicated by the vertical dashed lines) and the value of $\Delta_{\rm GH}$ at   $\omega^{\rm GH}_{\theta_{\rm m}}$ depend on $\theta_{\rm m}$. Again an oscillatory behavior with the Moir\'e angle is observed and the numerical data can be fitted using the functions $f(\theta_{\rm m})$ and $g(\theta_{\rm m})$ introduced previously (see caption for the value of fitting parameters). 

Finally, we would like to draw attention to the fact that Eqs. (\ref{DeltarApprox}) and (\ref{DeltaGHApprox}) can be inverted in order to obtain the bilayer's conductivity in terms of $\Delta_{\rm r}^{\theta_{\rm i}= \theta_{\rm B}}$ and $\Delta_{\rm GH}^{\theta_{\rm i}= \theta_{\rm B}}$, 
\begin{equation}
\sigma = \frac{2(n^2-1)}{n\sqrt{n^2+1}k_0Z_0}\frac{1}{\Delta_{\rm r}-\Delta_{\rm r}^{\rm subs}-i\frac{2n^2}{(n^2+1)^2}\Delta_{\rm GH}}\, ,
\end{equation}
where we defined $\Delta_{\rm r}^{\rm subs} = [\frac{n^2+3}{n^2+1}-1]/nk_0$ and omitted the superscript $\theta_{\rm i}= \theta_{\rm B}$ for clarity. In other words, the conductivity of bilayer graphene superlattices can be directly measured via the SHEL of light and spatial GH shift for a $p$-polarized gaussian beam impinging on the system at the Brewster angle of the substrate. These shifts can be easily accessed experimentally since they exceed the wavelength of incident light by a few orders of magnitude, making sub-wavelength photonic interactions a suitable and powerful tool for far-field characterization of electronic properties of two-dimensional materials. 

\section{Conclusion}

In summary, we have investigated the interplay between the photonic spin Hall effect and van Hove singularities in TBG. We demonstrated that a gaussian beam acquires a spin texture upon reflection at a bilayer graphene coated interface and it carries clear fingerprints of the Moir\'e pattern in the superlattice. The resulting SHEL shifts associated to left and right polarization states are several orders of magnitude larger than the wavelength of incident light, which is well within current experimental capabilities for detecting light beam shifts in the visible and near-infrared ranges \cite{Hosten2008,Luo2011Brewster, Herrera2010,Zhou20122,Qiu2014,Menard2009,Menard2010} .  The dispersion curves of the SHEL continuously move as the rotation angle in the bilayer is changed, allowing us to accurately identify the Moir\'e angle in the system. Our proposed models to fit the resonance's position as a function of the rotation angle are simple enough to guide experimental developments in the near future. We also showed that when combined to the Goos-H\"anchen effect, the SHEL enables full characterization of the electronic conductivity of  bilayer graphene via far-field optical measurements.  Altogether, our results demonstrate that spin-orbit interactions of light  are an appealing toolset for metrology of multilayer two-dimensional materials.   

\section*{Acknowledgement}
The authors are grateful to D. Dalvit, C. Farina, C. Matos, and F. Rosa for fruitful discussions. We acknowledge the use of supercomputing facilities at High Performance Computing Center (NACAD), COPPE, UFRJ. W.K.K. acknowledges the Los Alamos National Laboratory (LANL) Laboratory Directed Research and Development (LDRD) program as well as the Center for Nonlinear Studies  (CNLS) for financial support. F.J.C., R.B.C., and F. A. P.  acknowledge CNPq, CAPES, and FAPERJ for financial support. F.A.P. also thanks the The Royal Society-Newton Advanced Fellowship (Grant no. NA150208) for financial support.

\end{document}